\newcommand{\astt}{AST}
\newcommand{\wire}{\emph{WIRE}}
\newcommand{\smei}{{SMEI}}
\newcommand{\coriolis}{\emph{Coriolis}}

\documentclass[
    ,final            
  ]
  {aipproc}

\layoutstyle{6x9}

\begin{document}

\title{Welcome back, Polaris the Cepheid\thanks{Presented as a 
poster by A.\ J.\ Penny at the Cool Stars 15 meeting
at St Andrews.}}




\author{H.\ Bruntt}{
  address={Institute of Astronomy, School of Physics A29, 
University of Sydney, NSW 2006, Australia}
}

\author{A.\ J.\ Penny}{
  address={School of Physics \& Astronomy, 
University of St Andrews, North Haugh, KY16, 9SS, UK}
}

\author{D.\ Stello}{
  address={Institute of Astronomy, School of Physics A29, 
University of Sydney, NSW 2006, Australia}
}

\author{N.\ R.\ Evans}{
  address={Smithsonian Astrophysical Observatory, 60 Garden St.,
Cambridge, MA 02138, USA}
}

\author{J.\ A.\ Eaton}{
  address={Center for Excellence in Information Systems, Tennessee
State Univ., Nashville, TN 37209, USA}
}

\begin{abstract}

For about 100 years the amplitude of the 4\,d pulsation in Polaris has decreased.
We present new results showing a significant increase in the amplitude 
based on $4.5$ years of continuous monitoring from the ground and with two satellite missions.

\end{abstract}

\maketitle


\section{Introduction}

Polaris is the brightest Cepheid in the sky with a single known 
pulsation mode with a period close to 4 days. 
The star was found to be variable
around 150 years ago \cite{schmidt1857} and the period was first 
identified about 100 years ago \citep{hertz1911}.
For the past century the pulsation amplitude has 
decreased from $\sim100$ to 20 mmag in $V$ (peak-to-peak) 
and the period has increased by 4.5\,s yr$^{-1}$ \citep{turner05, ferro83}.

If the decrease in amplitude had continued, 
a possible explanation offered for the decrease in amplitude 
is that Polaris has completed its evolution through the
instability strip and is therefore becoming stable \cite{dinshaw89, kamper98}.
The high rate of period change is typical of overtone 
pulsators \cite{evans02} like Polaris, though other explanations 
have been suggested \cite{turner06, spreckley08}.


About fifteen years ago a paper with the title
\emph{Goodbye Polaris the Cepheid} \cite{fernie93}
predicted that the 4\,d mode would disappear around 1994.
However, a fundamental error was found in that paper \citep{kamper98},
and new measurements at the end of millennium indicated that the decrease
in amplitude had ceased \citep{kamper98, hatzes00}. 
Coincidentally, in 2008 three independent groups 
have reported unambiguous evidence for
a significant increase in the amplitude
\citep{spreckley08, bruntt08, lee08}, 
and we can now say, \emph{Welcome back, Polaris!}

Details of our work are given in \cite{bruntt08};
here we shall summarize our most important conclusions
and make comparisons with the results from the two other groups.
In addition we shall discuss a possible explanation for the surprising
increase in the amplitude of Polaris, 
based on new results for Cepheids in the Large Magellanic Cloud \cite{moska08}.

 \begin{figure}
  \includegraphics[height=.38\textheight]{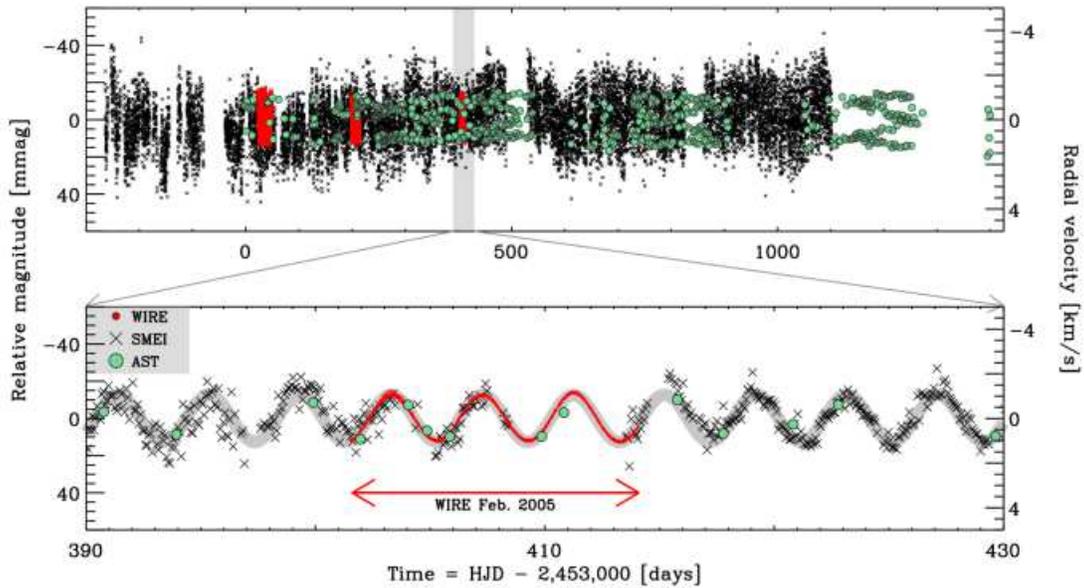}
  \caption{Light curves of Polaris from \wire\ and \smei\ and 
radial velocities from \astt. The top panel is the
entire data set with a baseline of 4.5 years;
the bottom panel shows the pulsation over 40 days.}
  \label{fig:lc}
\end{figure}

\section{Observations from ground and space}

We have collected extensive data for Polaris from
both ground and space. We used the 2-m Tennessee
State University Automatic Spectroscopic Telescope
(\astt\ \cite{eaton07}) to collect over 500 spectra over
a period of 3.8 years from 2003--7. 
We used the 52\,mm star tracker on the \wire\ satellite \citep{bruntt07wire} 
to collect 62 days of 
high-precision photometry during three runs, 
each lasting from 2 to 4 weeks, in 2004--5. Finally, we obtained
photometry from the \smei\ instrument on the \coriolis\
satellite spanning 3.7 years from 2003--6.


\section{A significant increase in the amplitude}

\begin{figure}
  \includegraphics[height=.35\textwidth]{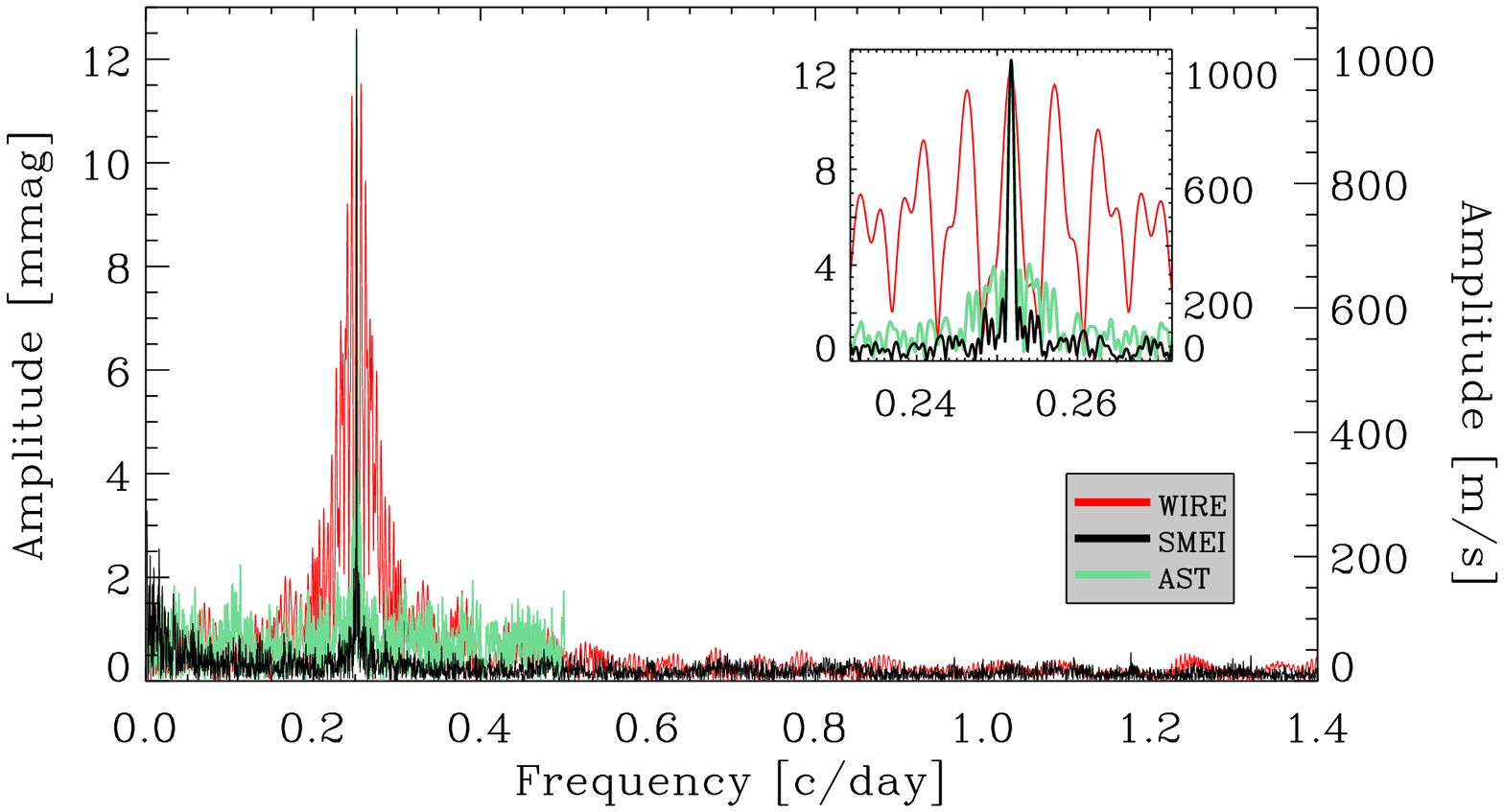}
 \hskip 0.5cm
  \includegraphics[height=.35\textwidth]{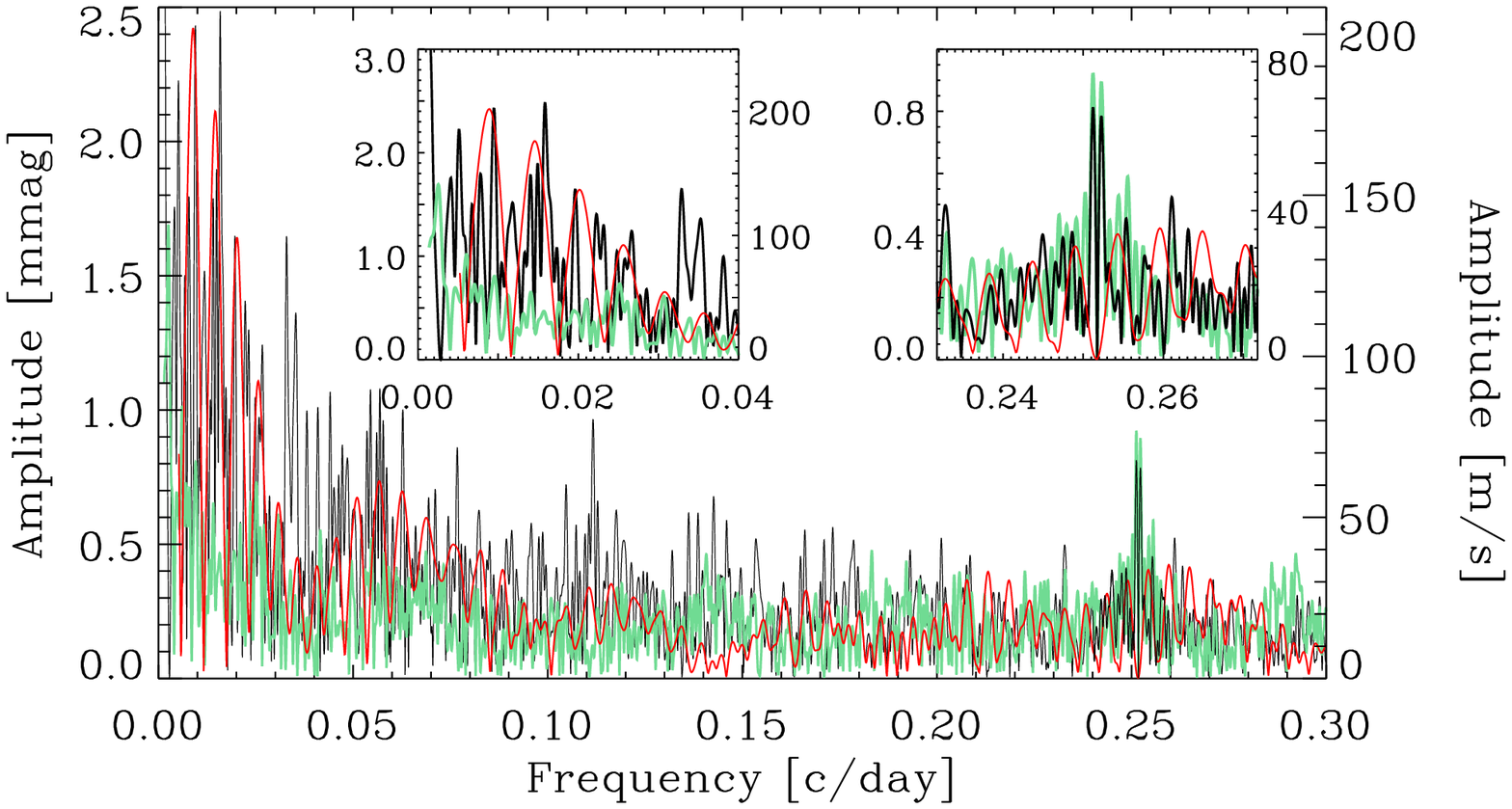}
   \caption{Left panel: amplitude spectrum of the three data sets. Right panel:
details of the amplitude at low frequencies after having subtracted the main 4\,d mode.}
  \label{fig:amp}
\end{figure}

In Fig.~\ref{fig:lc} we show the light curves from \smei\
and \wire\ and the radial velocity curve from \astt. 
The baseline of the combined data set is 4.5 years. 
To analyse the data we compute the Fourier amplitude
spectrum of each data set as shown in Fig.~\ref{fig:amp}.
The left panel shows the dominant peak due to the 4\,d pulsation mode. 
The right panel shows the details at lower frequencies
after having subtracted the primary mode.
The insets show details around the main features in the amplitude spectra.
Notice that in the right panel there is a significant residual
signal around the main peak at $\simeq0.25$ c/day. 
This is due to an increase in amplitude during 
the time span of the observations \cite{bruntt08}.

\begin{figure}
  \includegraphics[height=.37\textwidth]{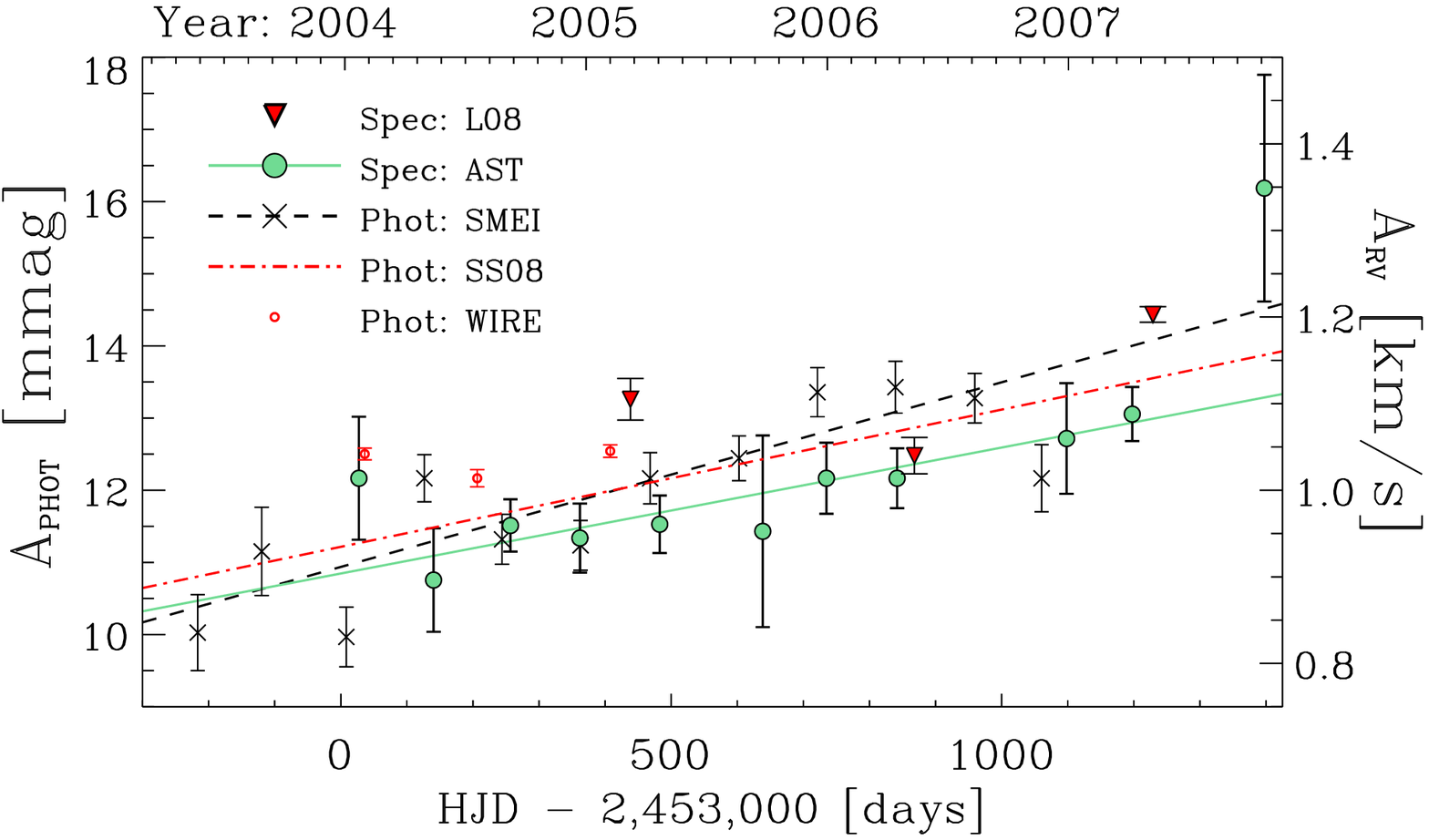}
\hskip 0.6cm
  \includegraphics[height=.37\textwidth]{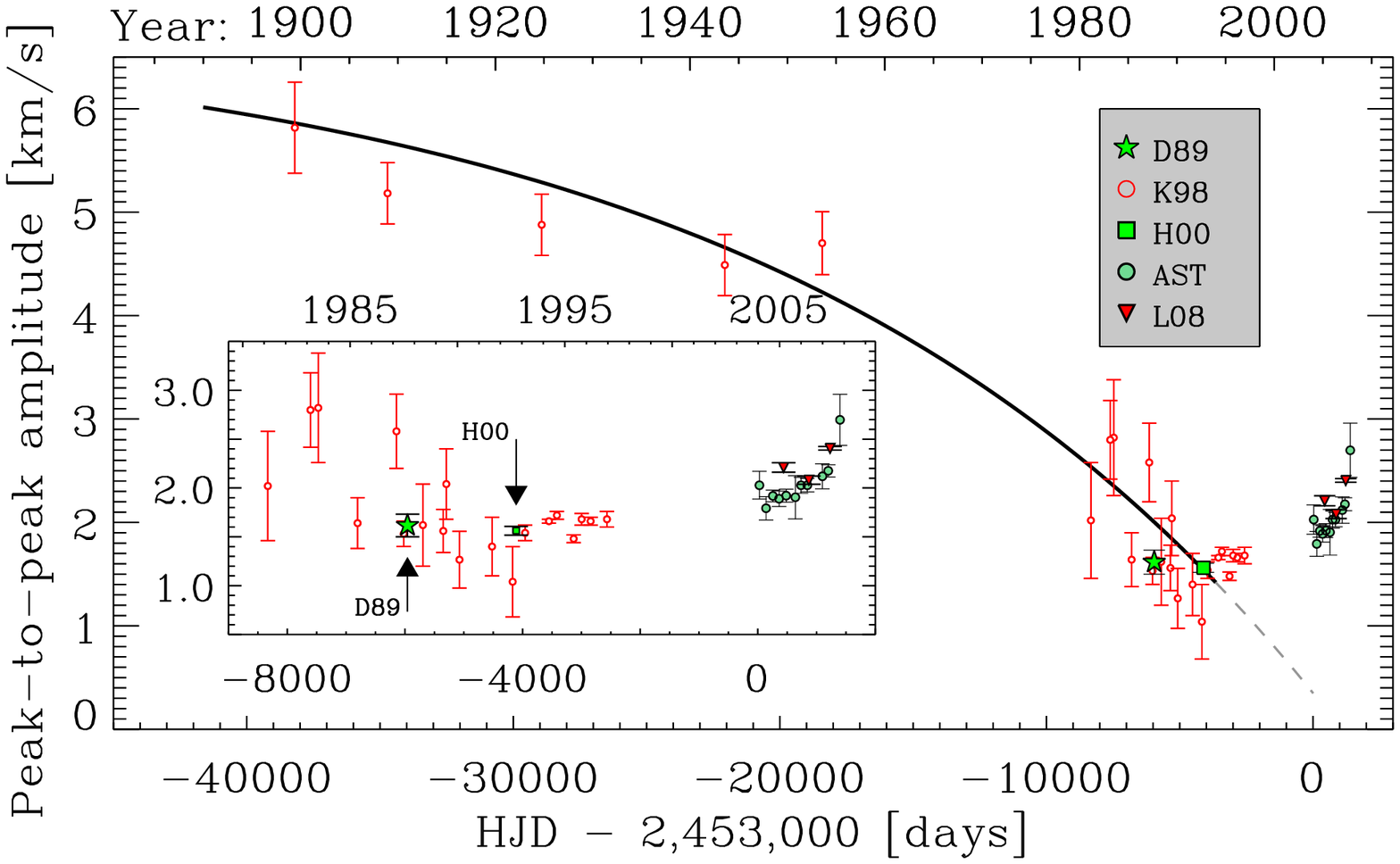}
   \caption{The left panel shows the measured increase in amplitude from out \astt\ and \smei\ data sets. 
The results from L08 \cite{lee08} and SS08 \cite{spreckley08} is also shown.
The right panel shows the change in radial velocity amplitude in Polaris over a century.}
  \label{fig:ampinc}
\end{figure}

To quantify the increase in amplitude, 
we split the data sets in groups of 30 pulsation cycles (4 months)
and measured the amplitude in each. 
The result is given in the left panel in Fig.~\ref{fig:ampinc}
showing a $\simeq30$\% increase in both photometric and radial-velocity amplitude.
In the right panel we show the change in radial-velocity amplitude over
a century and note that the precision of the data collected since the 
1990s have improved greatly with the advent of very stable spectrographs. 
We can see that the amplitude of the 4\,d mode in Polaris
stopped its decrease around 1995 and has increased since then.

Our measurement of the increase in amplitude 
is in very good agreement with the results from \cite{spreckley08}
and \cite{lee08} which are also shown in Fig.~\ref{fig:ampinc} 
(marked as SS08 and L08, respectively).
SS08 used an independent reduction pipeline
to reduce the \smei\ data and obtained 
a significantly better point-to-point precision.
L08 used 264 spectra collected over 2.6 years and
partly overlapping with our dataset.

\section{Long-period variation in Polaris}

In addition to the dominant 4\,d mode,
long-period variation has been reported in some of 
the extensive campaigns done on Polaris since the 1980s. 
These include variation at
$P=9.75$\,d ($A_{\rm rv}\simeq 1$\,km/s) \cite{kamper84},   
$45.3$\,d ($505\pm45$\,m/s) \cite{dinshaw89},  
$17.2$\,d and $40.2$\,d ($\simeq200$ m/s) \cite{hatzes00}, 
and most recently at 119.1\,d ($138\pm8$\,m/s) \cite{lee08},  
with semi-amplitudes given in the parentheses.
Based on simulations \cite{bruntt08} we 
argue that these modes are probably not intrinsic to Polaris,
but due to a combination of 
instrumental drift, 
too sparse sampling of the pulsation, 
and difficulties imposed by the period being so close to an integer number of days.
Our spectroscopic data from \astt\
has the lowest noise at low frequencies among the published data sets.
In the range 3--50\,d days we set an upper limit for the
variation at 100\,m\,s$^{-1}$, which is four times the average 
noise level in the amplitude spectrum. 


\section{Why has the amplitude increased?}

The increase in the amplitude of Polaris makes it unlikely that it is 
evolving out of the instability strip and ceasing to pulsate.  
Rather, the amplitude change seems to be a cyclic phenomenon, related to pulsation.
Based on comparison with pulsation models \cite{spreckley08} 
suggested that Polaris is on the red edge of the instability
strip for overtone pulsators, and that it could be evolving 
(depending on its mass) from a first-overtone pulsator to become 
a fundamental or double-mode Cepheid (also suggested by \cite{kamper98}).

Another potential explanation is beating of the primary mode 
with an additional unresolved pulsation mode,
as suggested by \cite{bruntt08} and \cite{spreckley08}.
If this be the case, the two modes would have nearly the same frequency, 
since the beat-period is at least 200 years. 
Recently, non-radial pulsation was detected in 42 out of 462 first-overtone Cepheids
in the Large Magellanic Cloud \cite{moska08}.
Interestingly, the non-radial frequencies are most often
located close to the primary radial overtone mode. 
These secondary peaks were not found in any of the 718 fundamental mode Cepheids \cite{moska08}.
Since Polaris is a first overtone pulsator \cite{feast97},
beating of the first overtone with a non-radial mode could explain the observed increase in amplitude.
However, this requires that the amplitudes of the two modes be comparable 
due to the low amplitude of the variation around the millennium. 
This is not the case for any of the LMC overtone Cepheids; 
the non-radial modes typically have amplitudes around 5\% of the main mode, 
with one Cepheid having a secondary peak with 20\% of the amplitude of the main mode.


\section{Outlook}

Fernie, Kamper \& Seager were apparently too pessimistic when
they said \emph{goodbye} to Polaris 10 years ago \cite{fernie93}.
Polaris has now evidently come back and is presenting
us with a new puzzle: why is the amplitude increasing?
In the coming years we will continue to monitor 
Polaris with \astt, and we will collect additional 
spectra at different longitudes to improve the spectral 
window of the observations.

\newpage

\bibliographystyle{aipproc}   

\bibliography{ms.bib}


\end{document}